%% file: paper_hepex_etaPlusMinus.tex
\documentclass{elsart} 
\journal{Physics Letters {\bf B}}
\usepackage{amssymb} 
\usepackage{graphicx} 
\usepackage{epsfig}
\usepackage[dvips]{color}

\begin{document}
\begin{frontmatter}
\vspace{0cm}
EUROPEAN ORGANIZATION FOR NUCLEAR RESEARCH\\
\vspace{0.7cm}
\begin{flushleft}
\hspace{9.5cm}\mbox{CERN-PH-EP/2006-034} \\
\hspace{9.5cm}\mbox{30 October 2006} \\ 
\end{flushleft}
\vspace{0.3cm}
\title{Measurement of the Ratio \mbox{\boldmath $\Gamma(K_L \rightarrow
    \pi^+\pi^-) / \Gamma(K_L \rightarrow \pi^{\pm}e^{\mp}\nu)$ and Extraction}
    \mbox{of the CP Violation Parameter \boldmath $|\eta_{+-}|$}} 
\date{} \input{authorlist}
\begin{abstract}
  We present a measurement of the ratio of the decay rates
  $\Gamma (K_L \rightarrow\pi^+\pi^-) / \Gamma(K_L
  \rightarrow\pi^{\pm}e^{\mp}\nu)$, denoted as
  $\Gamma_{K2\pi}/\Gamma_{Ke3}$. The analysis is based on data taken
  during a dedicated run in 1999 by the NA48 experiment at the CERN
  SPS. Using a sample of 47000 $K_{2\pi}$ and five million $K_{e3}$
  decays, we find $\Gamma_{K2\pi}/\Gamma_{Ke3} = (4.835 \pm
  0.022_{stat.} \pm 0.016_{syst.}) \times 10^{-3}$. From this we
  derive the branching ratio of the CP violating decay $K_L
  \rightarrow\pi^+\pi^-$ and the CP violation parameter
  $|\eta_{+-}|$. Excluding the CP conserving direct photon emission
  component $K_L \rightarrow \pi^+\pi^-\gamma$, we obtain the results
  BR$(K_L \rightarrow \pi^+\pi^-) = (1.941 \pm 0.019) \times 10^{-3}$
  and $|\eta_{+-}| = (2.223 \pm 0.012) \times 10^{-3}$.
\end{abstract}

\end{frontmatter}
\setcounter{footnote}{0}

%
%

\newcommand{\vet}{\protect\varepsilon_{\theta\theta}}
\newcommand{\vez}{\protect\varepsilon_{zz}} \pagestyle{headings}


\renewcommand{\topfraction}{1.}  \renewcommand{\bottomfraction}{1.}
\renewcommand{\textfraction}{0.}  \renewcommand{\floatpagefraction}{1.}

\newcommand{\egKK}{e.g.}  \newcommand{\eKK}{\mbox{$e$}}
\newcommand{\mrm}{\mathrm} \newcommand{\bfS}{{\bf{S}}}
\newcommand{\CP}{\mbox{$\mathrm{CP}$}}
\newcommand{\eps}{\mbox{$\epsilon$}}
\newcommand{\epsp}{\mbox{$\epsilon'$}}
\newcommand{\epsbar}{\mbox{$\bar{\epsilon}$}}
\newcommand{\eoe}{\mbox{$\re(\epsp/\eps)$}}
\newcommand{\ese}{\mbox{$\epsp/\eps$}}
\newcommand{\kz}{\mbox{$\mathrm{K^0}$}}
\newcommand{\kl}{\mbox{$K_L$}}
\newcommand{\kzbar}{\mbox{$\mathrm{\overline{K^0}}$}}
\newcommand{\ket}{\mbox{$K_{e3}$}}
\newcommand{\kmut}{\mbox{$K_{\mu 3}$}}
\newcommand{\ktpi}{\mbox{$K_{2\pi}$}}
\newcommand{\kthreepi}{\mbox{$K_{3\pi}$}}
\newcommand{\psum}{\mbox{$P_{sum}$}}
\newcommand{\ratio}{\mbox{$\Gamma_{K2\pi}/\Gamma_{Ke3}$}}

\newcommand{\kpit}{\mbox{$\mathrm K_{\pi3}$}}
\newcommand{\ks}{\mbox{$K_S$}}
\newcommand{\ksl}{\mbox{$\mathrm{K_{S,L}}$}}
\newcommand{\pim}{\mbox{$\mrm\pi^{-}$}}
\newcommand{\pip}{\mbox{$\mrm\pi^{+}$}}
\newcommand{\piz}{\mbox{$\mrm\pi^{0}$}}
\newcommand{\rell}{\mbox{$\mathrm R_{\mathrm ell}$}}
\newcommand{\pitomuv}{\mbox{$W(\pi\rightarrow MUV)$}}
\newcommand{\pitomuvData}{\mbox{$W_{Data}(\pi\rightarrow MUV)$}}
\newcommand{\pitomuvMC}{\mbox{$W_{MC}(\pi\rightarrow MUV)$}}
\newcommand{\kstocpipi}{\mbox{$\ks \rightarrow\pi^+\pi^-$}}
\newcommand{\kltocpipi}{\mbox{$\kl \rightarrow\pi^+\pi^-$}}
\newcommand{\ktocpipi}{\mbox{$\mathrm{K}\rightarrow\pi^+\pi^-$}}
\newcommand{\kstonpipi}{\mbox{$\ks \rightarrow\pi^0\pi^0$}}
\newcommand{\kltonpipi}{\mbox{$\kl \rightarrow\pi^0\pi^0$}}
\newcommand{\kltontpi}{\mbox{$\kl \rightarrow 3 \pi^0$}}
\newcommand{\ktonpipi}{\mbox{$\mathrm{K}\rightarrow\pi^0\pi^0$}}
\newcommand{\kltotpi}{\mbox{$\kl \rightarrow\pi^+\pi^-\pi^0$}}
\newcommand{\im}{\mbox{$\mathcal{I}m$}}
\newcommand{\re}{\mbox{$\mathcal{R}e$}}
\newcommand{\isoxi}[1]{\mbox{$\frac{\im A_{#1}}{\re A_{#1}}$}}
\newcommand{\rea}[1]{\mbox{$\re A_{#1}$}}
\newcommand{\ima}[1]{\mbox{$\im A_{#1}$}}
\newcommand{\minus}[1]{\mbox{$-#1$}} 
\newcommand{\ketKK}[2]{$|#1 \ #2 \rangle$} 
\newcommand{\etapm}{\mbox{$|\eta_{+-}|$}}
\newcommand{\etazz}{\mbox{$\eta_{00}$}}
\newcommand{\XPT}{\mbox{$\mrm{\chi PT}$} }
\newcommand{\als}{\mbox{$\alpha_{LS}$}}
\newcommand{\asl}{\mbox{$\alpha_{SL}$}}
\newcommand{\alspm}{\mbox{$\alpha_{LS}^{+-}$}}
\newcommand{\aslpm}{\mbox{$\alpha_{SL}^{+-}$}}
\newcommand{\alszz}{\mbox{$\alpha_{LS}^{00}$}}
\newcommand{\aslzz}{\mbox{$\alpha_{SL}^{00}$}}
\newcommand{\dals}{\mbox{$\Delta\als$}}
\newcommand{\dasl}{\mbox{$\Delta\asl$}}
\newcommand{\ptprime}{\mbox{$p_{\mathrm{t}}'$}}
\newcommand{\ptprimesq}{\mbox{$p_{\mathrm{t}}'^2$}}
\newcommand{\tten}[1]{\mbox{$\times 10^{#1}$}}

\newcommand{\laba}[1]{\label{sec:#1}}
\newcommand{\labc}[1]{\label{sec:#1}}
\newcommand{\labe}[1]{\label{equ:#1}}
\newcommand{\labs}[1]{\label{sec:#1}}
\newcommand{\labf}[1]{\label{fig:#1}}
\newcommand{\labt}[1]{\label{tab:#1}}
\newcommand{\refa}[1]{\ref{sec:#1}}
\newcommand{\refc}[1]{\ref{sec:#1}}
\newcommand{\refe}[1]{\ref{equ:#1}}
\newcommand{\refs}[1]{\ref{sec:#1}}
\newcommand{\reff}[1]{\ref{fig:#1}}
\newcommand{\reft}[1]{\ref{tab:#1}} \newcommand{\eq}[1]{(\refe{#1})}
\newcommand{\Eq}[1]{Gleichung~\refe{#1}}
\newcommand{\Eqs}[1]{Eqs.~(\refe{#1})}
\newcommand{\Eqss}[2]{Eqs.~(\refe{#1}) and (\refe{#2})}
\newcommand{\Eqsss}[3]{Eqs.~(\refe{#1}), (\refe{#2}), and (\refe{#3})}
\newcommand{\Fig}[1]{Fig.~\reff{#1}}
\newcommand{\fig}[1]{fig.~\reff{#1}}
\newcommand{\Figs}[1]{Figs.~\reff{#1}}
\newcommand{\Figss}[2]{Figs.~\reff{#1} and \reff{#2}}
\newcommand{\Figsss}[3]{Figs.~\reff{#1}, \reff{#2}, and \reff{#3}}
\newcommand{\Section}[1]{Section~\refs{#1}}
\newcommand{\Anh}[1]{Anhang~\refa{#1}}
\newcommand{\Chap}[1]{Chapter~\refc{#1}}
\newcommand{\Sec}[1]{Sec.~\refs{#1}}
\newcommand{\Secs}[1]{Sects.~\refs{#1}}
\newcommand{\Secss}[2]{Abschnitte~\refs{#1} und \refs{#2}}
\newcommand{\Secsss}[3]{Sects.~\refs{#1}, \refs{#2}, and \refs{#3}}
\newcommand{\tab}[1]{table~\reft{#1}}
\newcommand{\Tab}[1]{Table~\reft{#1}}
\newcommand{\Tables}[1]{Tabellen~\reft{#1}}
\newcommand{\Tabless}[2]{Tabellen~\reft{#1} und \reft{#2}}
\newcommand{\Tablesss}[3]{Tabellen~\reft{#1}, \reft{#2}, and \reft{#3}}

%
%
\section{Introduction}
In the last two years, the present generation of high-statistics kaon
experiments (KTeV, KLOE and NA48) have published various measurements
of the main \kl~decay modes, several of them being in disagreement
with the PDG averages given in \cite{PDG}. In \cite{KTEV_eta+-},
results for the six largest \kl~branching fractions were presented,
determined by measuring ratios of decay rates, where the charged decay
modes were normalized to \ket. The measurement of the ratio
\ratio~disagrees with the PDG by $10\,\%$, and the results for
$BR(\kltocpipi)$ and \etapm~disagree with the PDG by $5\,\%$, or more
than four standard deviations, respectively.

The analysis of the data collected by NA48 can clarify the
situation. In \cite{na48_vus}, we reported on
the measurement of the ratio of \ket~to all \kl~decays with two
charged tracks, leading to a branching ratio $BR(\ket)$ which exceeds
the PDG value by ($3.3 \pm 1.3$)\,\%, or 2.5 standard deviations. The
analysis presented here is based on that measurement. We used the same
data sample and applied similar cuts to select events with two
tracks and \ket~decays. 
%
%
\section{Description of the Experiment}
%
%
\subsection{Apparatus}
The NA48 experiment at the CERN proton synchrotron SPS was originally
designed and used for the precision measurement of direct CP violation
in kaon decays.  The NA48 beam line, detector and event reconstruction
have been described in detail elsewhere \cite{Alai}. Here we give a brief
summary of the main components relevant for this measurement. It
was performed using data collected in 1999 in a pure $K_L$ beam,
which was produced by an extracted 450\,GeV/{\it c} proton
beam striking a beryllium target at an angle of 2.4\,mrad. The last of
three collimators, located 126\,m downstream  of the target, defined
the beginning of the decay region in a 90\,m long vacuum
tank. 

Following a thin kevlar window, a tank filled with helium near
atmospheric pressure contained the magnetic spectrometer to measure
the momenta of the charged particles. It consisted of four drift
chambers (DCH), each with 8 planes of sense wires oriented along four
directions, each one rotated by 45 degrees with respect to the
previous one. The spectrometer magnet was a dipole with a field
integral of 0.883\,Tm, and was placed after the first two
chambers. The distance between the first and the last chamber was 21.8
meters. The momentum resolution was given by $\sigma(p)/p = 0.48\,\%
\oplus 0.009 \cdot p\,\%$ ($p$ in GeV/{\it c}). The spatial resolution
achieved per projection was 100\,$\mu$m, and the time resolution for
an event was $\sim 0.7$\,ns.

The hodoscope was placed downstream from the last drift chamber. It
consisted of two planes of plastic scintillators segmented in horizontal and
vertical strips and arranged in four quadrants. The signals were used
for a fast coincidence in the trigger. The time resolution from the
hodoscope was $\sim 200$\,ps per track.

The electromagnetic calorimeter (LKr) was a quasi-homogeneous
liquid krypton ionization chamber. Thin Cu-Be ribbon electrodes,
extending from the front to the back of the detector in a small-angle
accordion geometry, formed longitudinal towers of about $2 \times
2$\,cm$^{2}$ cross section to divide the active volume into 13248
readout cells. The calorimeter was 27 radiation lengths long, and
fully contained electromagnetic showers with energies up to
100\,GeV. The energy resolution was $\sigma(E)/E =
3.2\,\%/\sqrt{E}\oplus 9.0\,\%/E \oplus 0.42\,\%$  ($E$ in GeV).

In order to distinguish between muons and pions, a MUon Veto (MUV)
system was installed as the final component of the NA48 detector. It
consisted of three planes of plastic scintillators, each shielded by an 80\,cm
thick iron wall, allowing only muons to pass and produce a signal in
the scintillators. The probability for a pion to penetrate the whole
detector was of the order of $10^{-3}$. The inefficiency of the system
was at the level of one permille, and the time resolution was below 1\,ns.
%
%
\subsection{\labs{dataSample}Data Sample and Monte Carlo Simulation}
The data sample used for this analysis was collected during a two-day
minimum bias run in September 1999, dedicated to study semileptonic
\kl~decays. The spectrometer magnet polarity was changed once, so that
about half of the statistics was taken with positive and negative
magnet current, respectively.

Charged decays were triggered by a two-level trigger system: The
first level $(L1)$ required two charged particles in the scintillator 
hodoscope. The second level trigger $(L2)$ used information from the
spectrometer, demanding a vertex defined by two tracks with opposite
charge. In addition, events requiring only the $L1$ condition were
recorded as control triggers with a downscaling of 20 to measure the
efficiency of the $L2$ trigger. A total of $\sim$80 million 2-track
events were recorded, reconstructed and subjected to offline
filtering. 
\pagebreak

To reproduce the detector response, a detailed GEANT\cite{Geant}-based
Monte Carlo (MC) simulation of the NA48 apparatus was employed.  The
MC includes event generation, radiative corrections, propagation of
particles through the detector and response of the different detector
elements. To account for radiative effects, we used the PHOTOS program
package \cite{Photos} to simulate inner bremsstrahlung ($IB$) in the
\kltocpipi\, decay mode. For \ket~decays, $IB$ was simulated using the
event generator KLOR \cite{KLOR}, a program which includes both real
photon emission and virtual exchange.  A total of 18 million
\ktpi~decays and 30 million \ket~decays were generated (within ranges
of the vertex position and kaon energy enlarged with respect to
the acceptance in the analysis). In order to match the data, half of
the MC sample was simulated for each magnet polarity. The simulated
events had to pass the same selection criteria as the data, described below.
%
%
\section{Data Analysis}
%
%
\subsection{\labs{eventSelection}Analysis Strategy}
The basic measurement of this analysis is the ratio $R$ = \ratio. After
defining a sample of good 2-track events, we separated the two decay
channels. To obtain a clean signal of the CP violating decay
\kltocpipi, we had to suppress the main \kl~decay modes by several
orders of magnitude, unavoidably rejecting also part of the
\pip\pim~decays. Inefficiencies of the event selection and signal
losses, which were not exactly reproduced by the MC simulation, had to
be measured precisely and corrected for. With the ratio \ratio~thus
obtained, we determined the branching ratio of the decay \kltocpipi
$$
BR(K_L \to \pi^+\pi^-) = \frac{\Gamma(K_L \to \pi^+\pi^-)}{\Gamma(K_L \to
    \pi e \nu)} \cdot BR(K_L \to \pi e \nu) 
$$
and the CP violation parameter \etapm
$$
|\eta_{+-}| \equiv \sqrt{\frac{\Gamma(K_L \to \pi^+\pi^-)}{\Gamma(K_S \to
\pi^+\pi^-)} } = \sqrt{\frac{BR(K_L \to \pi^+\pi^-)}{BR(K_S \to
	\pi^+\pi^-)} \cdot \frac{\tau_{KS}}{\tau_{KL}} }\,\,.
$$
Note that throughout the analysis, only information
from charged tracks was used. As a result, we accepted any number of
photons in the events, so that e.g. the radiative $K_L \to
\pi^+\pi^-\gamma$ decays were also accepted. The consequences for our
results are described in detail in \Sec{results}. 
%
%
\subsection{Basic 2-track Selection}
The sample of good 2-track events was extracted from the recorded
events by applying the following selection criteria: the events were
required to have exactly two tracks with opposite charge, meeting at a
distance of closest approach below 3\,cm to define the decay
position. This vertex had to be inside a cylinder 3\,cm in radius
around the beam axis and longitudinally between 8\,m and 33\,m from
the final collimator.

To guarantee good reconstruction efficiency, events with high hit
multiplicity in the drift chambers (i.e. more than seven hits in a
plane within 100\,ns) were rejected. Tracks were accepted within the
momentum range 15\,GeV/{\it c} to 100\,GeV/{\it c}, and their
extrapolations had to be within the geometrical acceptances of the
various subdetectors. The track times, given by the spectrometer, were
required to coincide, admitting a maximum difference of 6\,ns.  In
order to allow a clear separation of showers, we required the distance
between the entry points of the two tracks at the front face of the
electromagnetic calorimeter to be larger than 25\,cm. These cuts were
passed by $\sim$20 million events.
%
%
\subsection{$\pi^+\pi^-$ Selection}
Additional cuts were applied to extract the \kltocpipi~sample, where
the two semileptonic \kl~decays, \ket~and \kmut, are the dominant
background sources.

The decay channel \kltotpi~was completely removed by requiring the
2-track invariant mass $m_{\pi\pi}$ to be compatible with the kaon
mass: $0.48\,{\rm GeV/{\it c}}^2 < m_{\pi\pi} < 0.51\,{\rm GeV/{\it c}}^2$. The
cut on $m_{\pi\pi}$ also rejected most of direct emission
$\pi^+\pi^-\gamma$ decays.

The missing momentum carried away by the undetected neutrino in the
semi-leptonic decays is reflected in the transverse component $p_t$ of
the reconstructed kaon momentum $p_K$. The requirement $p_t^2 < 3\times
10^{-4}\,{\rm GeV}^2/c^2$ suppressed the semileptonic background
significantly. 

Further elimination of \ket~decays was achieved by requiring the ratio
$E/p$ for each track to be less than 0.93, where $E$ is the energy
deposited in the electromagnetic calorimeter, and $p$ is the track
momentum measured in the magnetic spectrometer. This cut was applied
only to the data.

Remaining background from \kmut~decays was strongly reduced due to the
high muon detection efficiency of the muon veto detector (better than
$99.9\,\%$). Events were rejected if a track with an associated
signal in the muon counters was found.
%
%
\subsection{$\pi^\pm e^\mp \nu$ Selection}
Being the only relevant \kl~decay channel with an electron in the
final state, \ket~events can be selected by applying only an $E/p$
criterion. If the ratio $E/p$ for any of the tracks exceeded 0.93, the
track was tagged as being due to an electron, thus classifying the
event as a \ket~decay. Like for the $\pi^+\pi^-$ selection, this cut
was applied only to the data, but not to the MC. The quantity $E/p$ is shown in
\Fig{eop_ke3} for the tracks of all selected $K_{e3}$ events.
\begin{figure}[ht]
\begin{center}
\includegraphics[width=10.5cm]{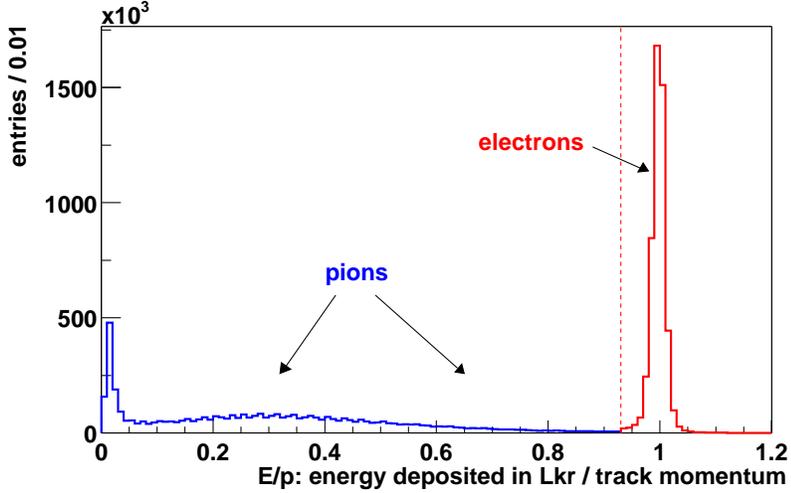}
\parbox{11cm}{\caption[] {\labf{eop_ke3}The ratio of calorimetric
  energy $E$ over the momentum $p$ for the tracks of all selected
  $K_{e3}$ events. The vertical dashed line at 0.93 indicates the default
  $E/p$ cut value.}}
\end{center}
\end{figure}

A final criterion was applied to a measure of the kaon momentum. We
required the sum of the track momenta \psum~to be between 60\,GeV/{\it c} and
120\,GeV/{\it c}, the boundaries being chosen to guarantee good
description by the MC and to have a negligible contribution from
\kstocpipi~(0.1 permille) due to \kl-\ks~interference.

After applying the cuts described above, 47142 \ktpi~and 4999126
\ket~candidates were selected from the data sample. Taking these event
numbers and correcting for the acceptances evaluated by MC and listed in
\Tab{DetectorAcceptancesForTheDecayModes}, the raw (uncorrected) value
for the ratio is $\ratio = (4.833 \pm 0.023) \times 10^{-3}$.
\begin{table}[ht]
\begin{center}
	\begin{tabular}[t]{|l|l|}\hline 
Decay mode & Acceptance\\ \hline  
$K_{2\pi}$ & $0.5826 \pm 0.0004$ \\
$K_{e3}$ & $0.2986 \pm 0.0002$ \\ \hline
	\end{tabular}
\parbox{11.5cm}{\caption[]
  {\labt{DetectorAcceptancesForTheDecayModes}Detector acceptances for
  events with 8\,m $<$ vertex position $<$ 33\,m and $60\,{\rm GeV/{\it c}}
  < \psum < 120\,{\rm GeV/{\it c}}$.}}
\end{center}
\end{table}
%
%
\section{\labs{systematics}Corrections and Systematic Uncertainties}
The following sections describe the determination of the corrections
and systematic uncertainties, generally measured from the data samples
themselves.
%
%
\subsection{\labs{corrMUV}Rejection of Events with Muon Signal}
Background from \kmut~decays in the \pip\pim~sample can only be
reduced to the desired level if we reject events having a track in
coincidence with an associated signal in the muon counters. A few
percent of the pions decay ($\pi\rightarrow\mu\nu$) or penetrate the
whole detector ({\it punch-through}), so we lose \ktpi~events cutting
on the MUV information. As decays after the LKr and pion punch-through
are not simulated in the MC, the probability \pitomuv~for a pion to
generate a hit in the MUV (directly or indirectly) is higher in data,
and we must correct for the difference $\Delta = \pitomuvData -
\pitomuvMC$.

As one cannot measure \pitomuv~for \ktpi~in data, the difference
$\Delta$ was determined from \kmut~candidates. We applied all
\ktpi~selection cuts, but required both tracks to have an associated
signal in the MUV. The remaining events had the same signature as the
\pip\pim~signal, being most probably \kmut~decays (with decay
of the pion), and representing a clean sample to study the MUV signal
probability for a pion. The result for this measurement is shown in
\Tab{piMUV_kmu3}. The numerator in data was corrected for the few
\ktpi~events with double pion decay. The $\Delta$ obtained was $\Delta =
(0.24 \pm 0.09)\,\%$, the uncertainty being completely dominated by the
statistical error.
\begin{table}[ht]
\begin{center}
\begin{tabular}{|c|c|c|}\hline
$\pitomuvData$ & $\pitomuvMC$ & $\Delta$ \\\hline $\frac{585}{40095} =
(1.46 \pm 0.06)\,\%$ & $\frac{358}{29413} = (1.22 \pm 0.07)\,\%$ &
$(0.24 \pm 0.09)\,\%$ \\\hline
\end{tabular}
\end{center}
\parbox{13cm}{\caption[] {\labt{piMUV_kmu3}\pitomuv~and the difference
$\Delta$ between data and MC, determined from \kmut~events surviving
the \pip\pim~selection.}}
\end{table} 

A number of alternative measurements were performed to verify the
$\Delta$ quoted above; as the additional amount of MUV signal in data
is mostly independent from the selection criteria, it is expected to
be similar in all control samples. \\
\kltotpi~decays were selected from the sample
of 2-track events, and 30 million \kltotpi~decays were simulated to
determine \pitomuv~in data and MC. We obtained $\Delta = (0.28 \pm
0.04)\,\%$.\\  
As a direct check, we measured \pitomuvData~from a data
sample taken in the year 2002 with a pure \ks~beam. We selected
\kstocpipi~decays applying the same cuts as for the
\kltocpipi~selection. With \pitomuvMC~from the \ktpi~MC, we measured
$\Delta = (0.23 \pm 0.01)\,\%$, which is in very good agreement with
the $\Delta$ measured from the \kmut~sample. As a result, the
correction on $R$ due to the MUV cut is $\Delta R\,(MUV) = 2 \times \Delta$
(two pions in the final state) $= (+0.48 \pm 0.18)\,\%$.
%
%
\subsection{\labs{corrTrig}Trigger Efficiency Differences}
As described in \Sec{dataSample}, the downscaled $L1$ trigger
sample was used to determine the $L2$ efficiency. We measured
$\epsilon_{L2}(\ktpi) = (99.76 \pm 0.10)\,\%$ and
$\epsilon_{L2}(\ket) = (98.47 \pm 0.02)\,\%$, resulting in a
correction due to trigger efficiency differences of 
$\Delta R\,(trigger) = (-1.29 \pm 0.11)\,\%$.
%
%
\subsection{\labs{corrEoverP}Cut on the Ratio $E/p$}
The purpose of the $E/p$ criterion is to distinguish between electrons
and pions. As the resulting separation is not definite, we had to
measure the misidentification probabilities precisely from the data,
and correct the numbers of \ktpi~and \ket~events accordingly. 
To account for the momentum spectra of the particles from different
decay channels, the determination of
the misidentification probabilities as well as the application of the
corrections were performed in bins of the track momentum.

The pion misidentification probability $W(\pi \rightarrow e)$ (pions
with $E/p \ge 0.93$, being classified as electrons) led to a loss of
selected \ktpi~decays, while background remained in the \ket~sample,
predominantly due to $K_{\mu 3}$. For the pion
misidentification measurement, a sample of \ket~events was selected
having one track with $E/p > 1.0$, clearly classifying it as an
electron. The probability for pions to have $E/p \ge 0.93$ was then
determined from the $E/p$ spectrum of the other (i.e. pion) track to
be $W(\pi \rightarrow e) = (0.592 \pm 0.006)\,\%$.

The inefficiency of the electron identification $W(e \rightarrow \pi)$
(electrons with $E/p < 0.93$, being classified as pions) reduced the
number of selected \ket~events. It was determined in a similar way by
requiring one track with $E/p < 0.7$, tagging it as a pion. The
$E/p$ distribution for the other track is then mainly due to
electrons, with a small contribution from pions. Subtracting this pion
component, we obtained the probability for losing an electron by the
condition $E/p \ge 0.93$: $W(e \rightarrow \pi) = (0.478 \pm
0.004)\,\%$.

The corrections were applied as follows: we increased the \ktpi~event
number by the factor $(1 + 2 \times W(\pi \rightarrow e))$ (two pions
in the final state). The \ket~number was increased by $(1 + W(e
\rightarrow \pi))$, and backgrounds from \kmut, \kthreepi~and
\ktpi~were subtracted. To evaluate the \kmut~background, we simulated
a sample of 30 million \kmut~MC events, applied the same cuts as for
the \ket~MC and determined the acceptance of this
selection. Normalized to the flux, the fraction of \kmut~events in the
\ket~data sample was about four permille. The background from
\kthreepi~(about two permille) and the small \ktpi~contribution
($\sim$0.1 permille) were determined in a corresponding way.

To demonstrate the correctness of the \ket~selection principle, we varied
the cut value between $E/p$ $> 0.85$ and
$E/p$ $> 0.96$, taking into account that the background in the
\ktpi~signal (see \Sec{corrBkgK2pi}) depends on the $E/p$ criterion as
well. The first line of \Tab{eopCutVariations2} shows the ratio $R$
before all corrections. The results for $E/p > 0.85$ and $E/p > 0.96$
differ by six percent, but the following $E/p$ dependent event number
corrections bring them to an excellent agreement. We derived the 
systematic uncertainty from the largest deviation of
$0.05\,\%$, and the correction due to the $E/p$ cut is $\Delta
R\,(E/p) = (+1.34 \pm 0.05)\,\%$, which includes inefficiencies of the
cut and background in \ket.
\begin{table}[ht]
  \begin{center}
    \begin{tabular}[t]{|l|c|c|c|}\hline
      \boldmath $ E/p$ \bf criterion & \bf 0.85 & \bf 0.93 & \bf 0.96 \\ \hline
      \color{blue}Raw $K_{2\pi}/K_{e3}$ [$10^{-3}$] & \color{blue}
      4.655 & \color{blue} 4.833 & \color{blue} 4.946 \\ \hline
      $E/p$ inefficiencies + \ket~background [$\%$]& $+\, 4.91$ & $+\, 1.34$ & $-\, 0.47$\\
      $K_{2\pi}$ background [$\%$]& $-\, 0.25$ & $-\, 0.49$ & $-\, 1.04$\\ \hline
      Total correction [$\%$] & $+\, 4.67$ & $+\, 0.85$ & $-\, 1.51$\\
      \color{red} Corrected $K_{2\pi}/K_{e3}$ [$10^{-3}$] &
      \color{red} 4.873 & \color{red} 4.874 & \color{red} 4.871\\ 
      Difference to standard ($E/p$ $> 0.93$) [$\%$] & $-\, 0.01$ & - & $-\, 0.05$\\ \hline
    \end{tabular}
  \end{center}
  \caption{Variation of the $E/p$ cut and its impact on the ratio
  \ratio.}
  \labt{eopCutVariations2}
\end{table}
%
%
\subsection{\labs{corrBkgK2pi}Background in the \ktpi~Signal}
Applying the \pip\pim~selection cuts, a suppression of the background
from \ket~and \kmut~of the order $10^4$ was
achieved. \Fig{k2pi_pipiMass} shows the distribution of the invariant
\pip\pim~mass $m_{\pi\pi}$ after all selection requirements except
the cut on $m_{\pi\pi}$ itself. The data are well described by the sum of the
\ktpi~signal MC and the two background MCs. The small mismatch between
the reconsructed and the true kaon mass in this particular data taking
period has not been corrected, as it has no effect on the selection
and the background estimation.

The total amount of background was determined from sidebands of the
distributions (sideband\,1: $0.40 - 0.45\,$GeV/$c^2$, sideband\,2:
$0.53 - 0.58\,$GeV/$c^2$). The combined background from \ket~and
\kmut~yields an estimation of $232 \pm 15$ events, leading to a
correction due to background in \ktpi~of  $\Delta R\,(bg \ktpi) =
(-0.49 \pm 0.03)\,\%$.
\begin{figure}[ht]
\begin{center}
\includegraphics[width=10.5cm]{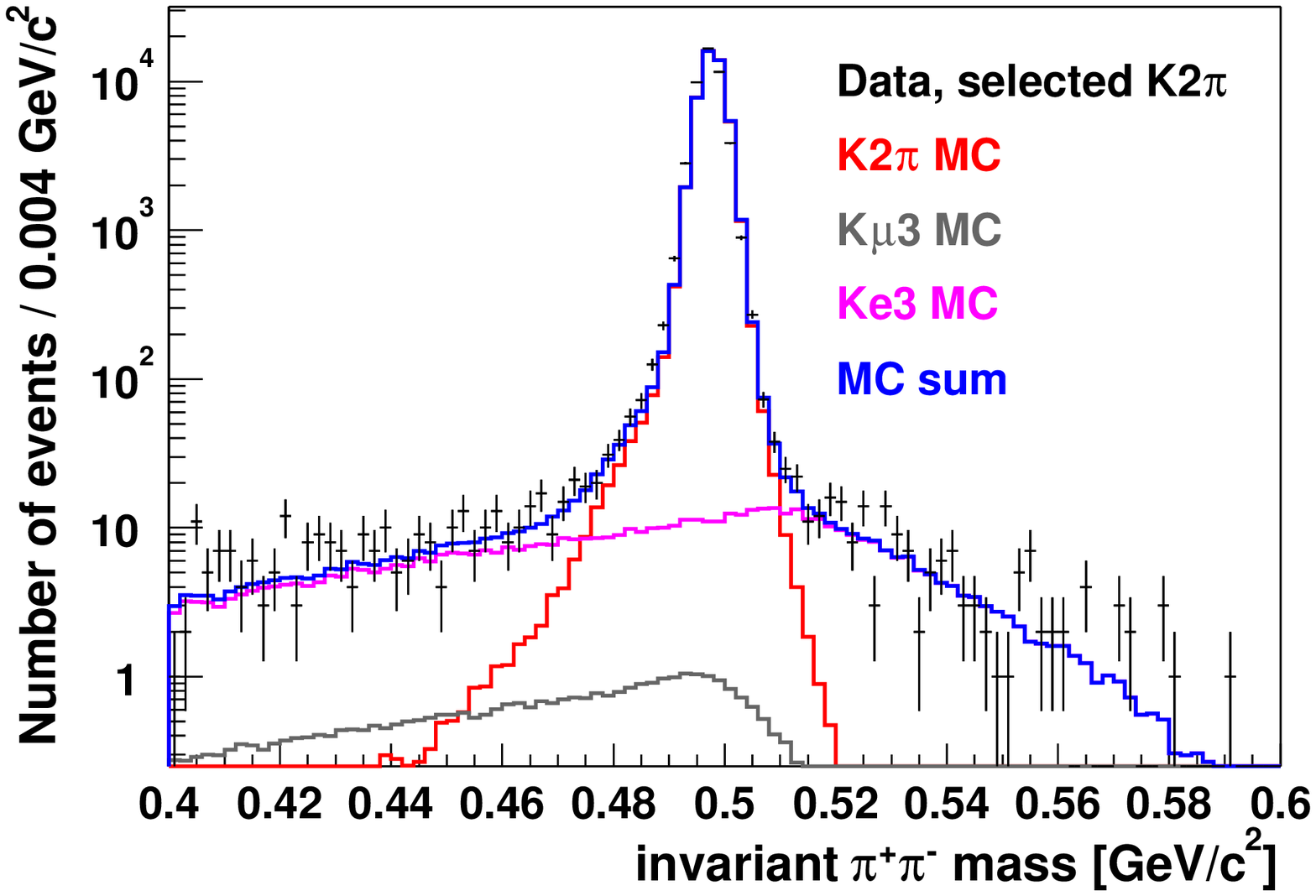}
\parbox{11cm}{\caption[] {\labf{k2pi_pipiMass}Distribution of the
  invariant \pip\pim~mass.}}
\end{center}
\end{figure}

As a check, we enlarged the accepted mass window
to $0.45 - 0.53\,$GeV/$c^2$, thus roughly doubling the background fraction.
Correcting for this, however, we obtained the same result for \ratio.
%
%
\subsection{\labs{spectrum}Kaon Energy Spectrum}
\begin{figure}[b]
\begin{center}
\includegraphics[width=10.cm]{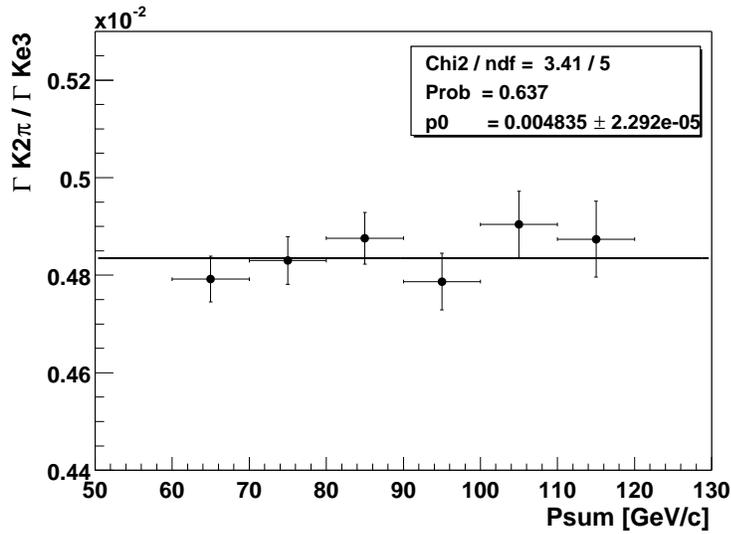}
\parbox{11cm}{\caption[] {\labf{ratio_psum}The ratio as a function of
  \psum. The errors are the combined uncertainties from data and MC statistics.}}
\end{center}
\end{figure}
To check for a dependence on the kaon momentum, we determined the
ratio $R$ in bins of \psum~(\Fig{ratio_psum}). Although the points
fit well to a straight line without slope, the significance of this
conclusion is affected by the relatively small \pip\pim~statistics. To disentangle the
\ktpi~contribution, we studied the $P_{sum}$ distributions for the
two decay modes separately. \Fig{psum} shows the \psum~spectra for
selected \ktpi~and \ket~events in data and MC. While there is no
visible slope in the ratio data/MC for \ktpi, the \ket~mode shows a small
dependence. In order to derive the spectrum error, we split the data
sample into halfs of the \psum~range, 60-90\,GeV/{\it c} and 90-120\,GeV/{\it c},
and compared the ratios data/MC in the two samples for the
\ket~mode. The difference between the mean ratios was taken as the
uncertainty due to the imperfect knowledge of the kaon energy spectrum: $\Delta
R/R\,(energySpectrum) = 0.20\,\%$.
\begin{figure}[ht]
\begin{center}
\includegraphics[width=13.cm]{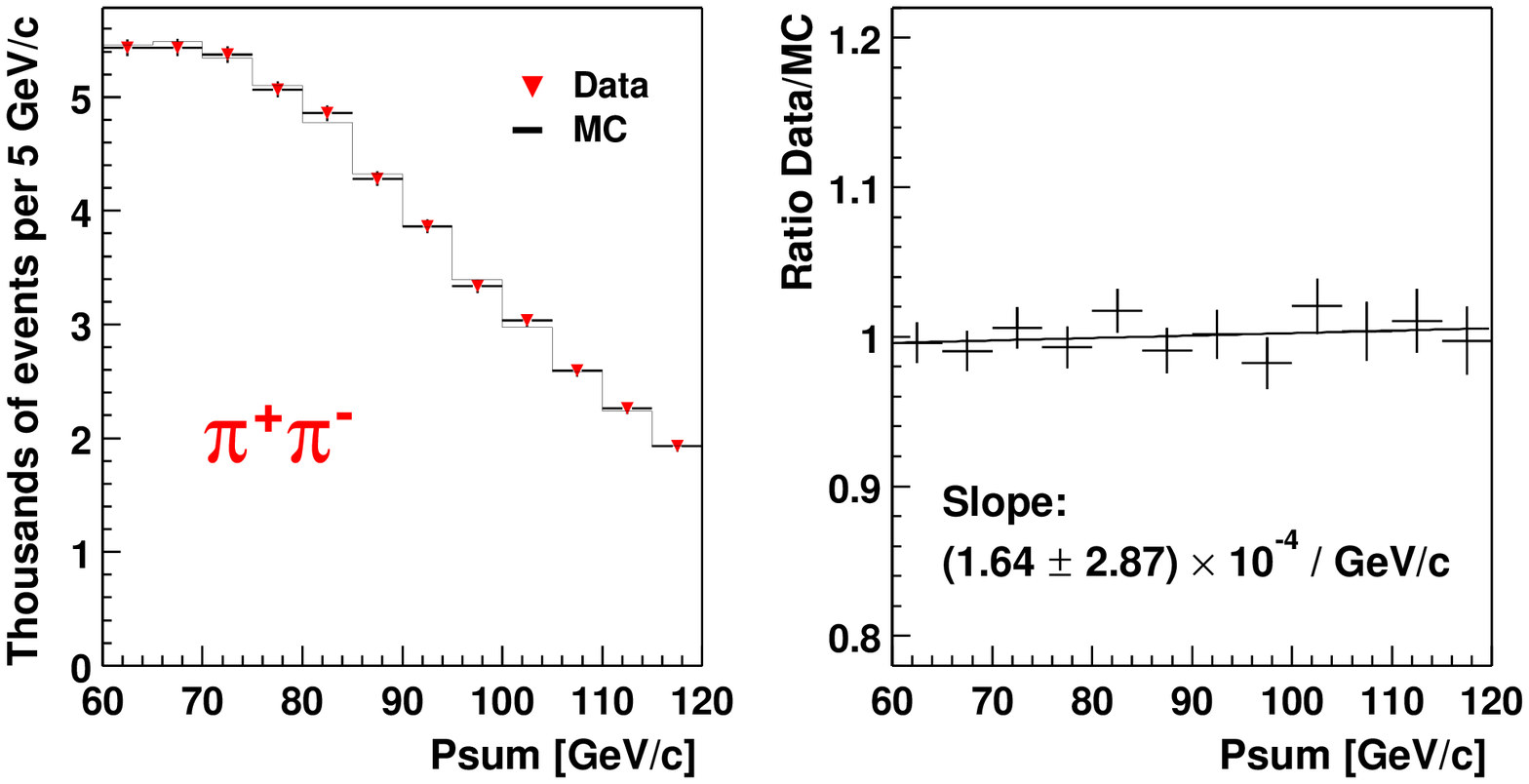}
\includegraphics[width=13.cm]{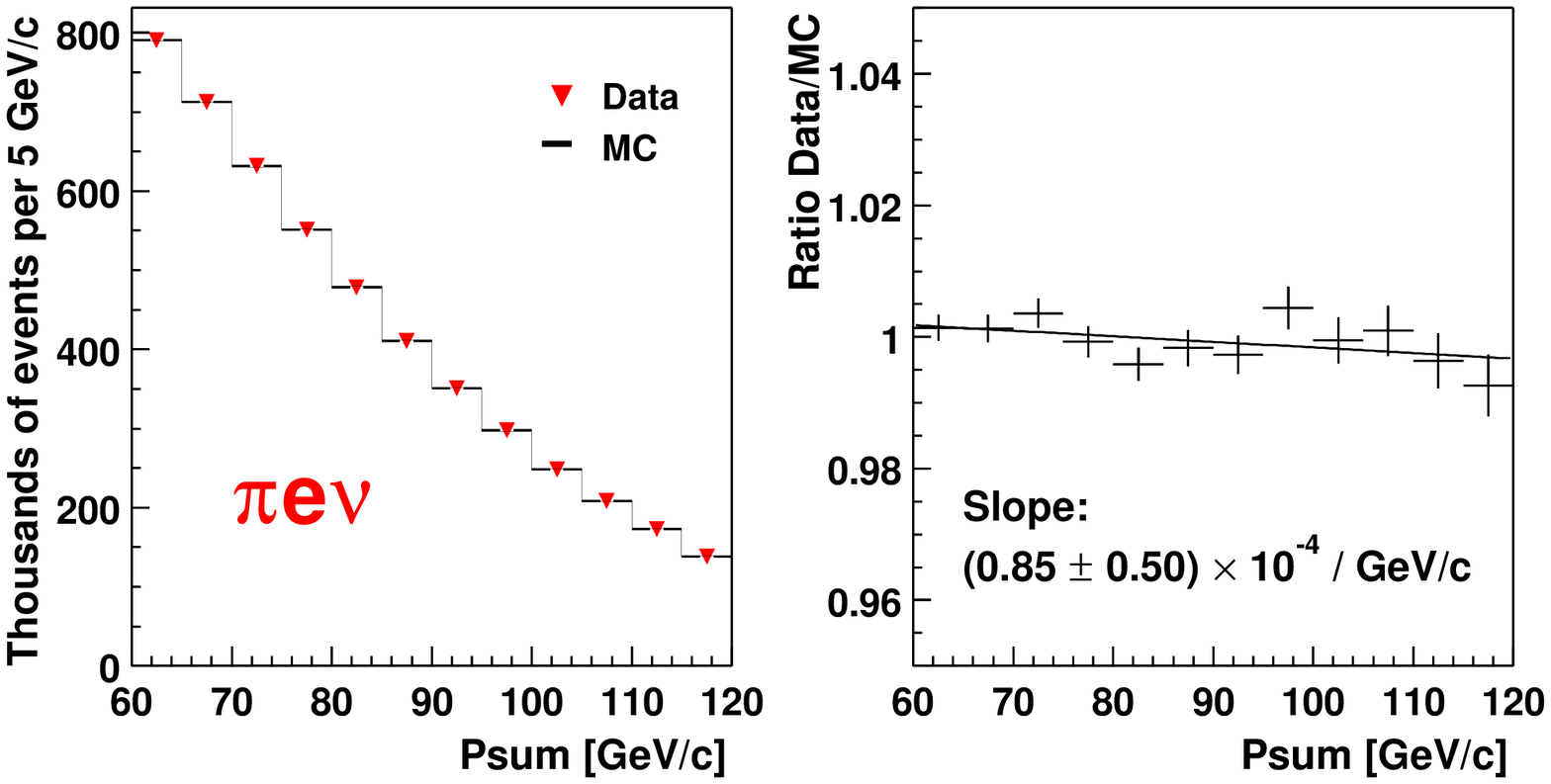}
\parbox{13.5cm}{\caption[] {\labf{psum}Sum of track momenta \psum~for
  selected \ktpi~(top) and \ket~(bottom) events. Left:
  distributions from data and MC. Right: ratios data over MC.}}
\end{center}
\end{figure}

As a check, we performed a number of variations of the accepted
\psum~range, e.g concentrating on a central region between 70 and
100\,GeV/{\it c}, enlarging the boundaries to 55-140\,GeV/{\it c}, or changing
only the lower or upper cut value. All
resulting deviations in $R$ were well below the estimated systematic
uncertainty.
%
%
\subsection{\labs{radcorr}Radiative Effects}
\begin{figure}[b]
\begin{center}
\includegraphics[width=13.4cm]{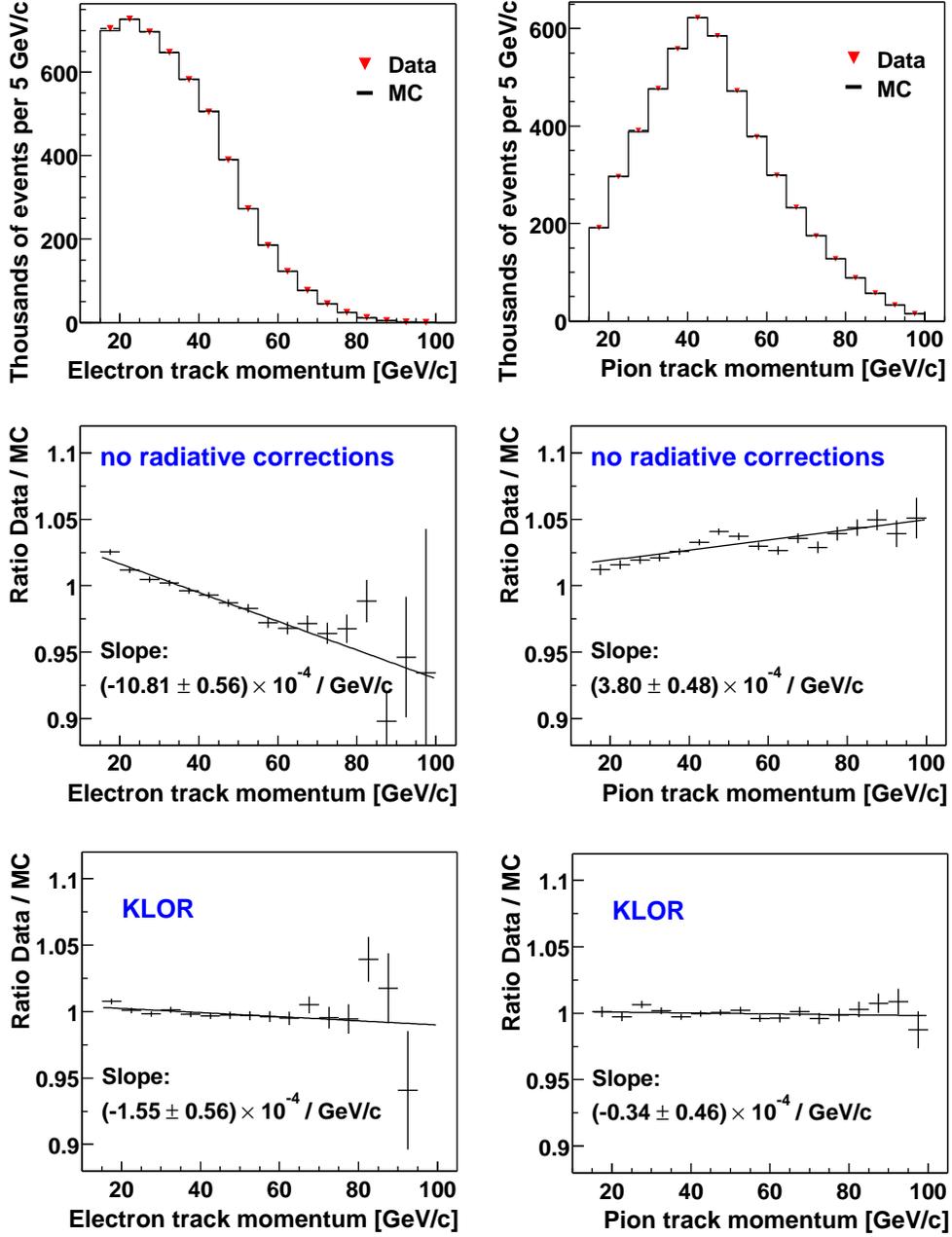}
\parbox{13.8cm}{\caption[] {\labf{ke3_ptrack_noradcorr}For selected
    \ket~events: Momentum spectra for the electron (top left) and pion
    (top right) in data and MC (using KLOR). Beneath follow the ratios
    data/MC: Without radiative corrections in the MC (middle), and with
    KLOR (bottom).}}
\end{center}
\end{figure}
As mentioned in \Sec{dataSample}, we used the programs PHOTOS (for
\ktpi) and KLOR (for \ket) to simulate inner bremsstrahlung.
The effect and importance of
the simulation is demonstrated in \Fig{ke3_ptrack_noradcorr}.
With KLOR, the momentum spectra of the charged particles are well
described.

To estimate the uncertainty from radiative effects, we
repeated the analysis using an updated version of the PHOTOS package,
now also including virtual diagrams in the bremsstrahlung
simulation. The results differed only by 0.1 permille, giving
confidence that the uncertainty is fully covered by $\Delta
R/R\,(radCorr) = 0.1\,\%$.
%
%
%
\section{\labs{results}Results}
%
%
%
\subsection{The Ratio \ratio}
We summarize the corrections and systematic uncertainties on $R$ in 
\Tab{systUncertainties}. 
\begin{table}[ht]
\begin{center}
\begin{tabular}{|l|r|r|}\hline
Source of uncertainty & Correction $[\%]$ & Uncertainty $[\%]$
\\\hline\hline
$E/p$ cut & $+\,1.34$ & 0.05 \\
Background in K$_{2\pi}$ & $-\,0.49$ & 0.03 \\
Muon cut & $+\,0.48$ & 0.18 \\
Trigger efficiencies & $-\,1.29$ & 0.11 \\
Energy spectrum & - & 0.20 \\
Radiative corrections & - & 0.10 \\\hline
MC statistics & - & 0.10 \\\hline\hline
Total correction & $+\,0.04$ & 0.33 
\\\hline
\end{tabular}
\parbox{11cm}{\caption[] {\labt{systUncertainties}Summary of relative
    corrections and systematic uncertainties on the ratio \ratio.}}
\end{center}
\end{table} 

Applying the small total correction of $+0.04\,\%$ to the raw ratio,
we obtain
\begin{eqnarray}
\frac{\Gamma (\kltocpipi)}{\Gamma(\kl\rightarrow\pi e \nu)} & = & 
(4.835 \pm 0.022_{stat.} \pm 0.016_{syst.}) \times 10^{-3}  \nonumber 
\\ & = & (4.835 \pm 0.027) \times 10^{-3}. \nonumber
\end{eqnarray}

A number of additional systematic checks were performed, varying some
of the selection cuts or dividing the data into subsamples to study the
result as a function of certain quantities, e.g. the longitudinal
vertex position or the magnet polarity. None of the variations showed
any significant impact on the result. 

As the uncertainty from the energy spectrum gives the largest
contribution to the systematic error, we performed an alternative
analysis, applying the same cuts but defining the kaon momentum for
\ket~decays in a different way. 
Both decay modes were required to have a kaon momentum $p_K$
between 70 and 140\,GeV/{\it c}. For $\pi^+\pi^-$ decays, $p_K$ is directly
given as the sum of the track momenta. For \ket~decays, however, the neutrino
leaves the detector undetected, leading to a quadratic ambiguity in
the determination of the kaon momentum. Depending on the orientation
of the longitudinal component of the neutrino momentum in the kaon
cms, there are two solutions for $p_K$. Both solutions were required
to be in the accepted momentum range. About 41500 \ktpi~and 2.66
million \ket~events passed the selection. The determinations of all
corrections and systematic uncertainties were repeated and performed
as described in the previous sections. After all corrections, we
obtained $\ratio = (4.835 \pm 0.024_{stat.} \pm 0.029_{syst.}) \times
10^{-3}$, which is in perfect agreement with our result. The larger
systematic error results from the ambiguous definition of the kaon momentum
for \ket~decays.
%
%
\subsection{The Branching Ratio BR(\kltocpipi)}
For the determination of $BR(\kltocpipi)$ and \etapm, we must consider
that the \pip\pim~event selection does not imply any requirement
concerning photons, i.e. the radiative decay $K_L \to
\pi^+\pi^-\gamma$ is also accepted, as long as the $\pi\pi$ invariant
mass fulfills the kaon mass requirement. $K_L \to \pi^+\pi^-\gamma$
decays can originate from two different processes; inner
bremsstrahlung ($IB$) or direct emission ($DE$). While the $IB$
process is CP violating (the photon is emitted by a pion, coming from
$K_L \to \pi^+\pi^-$), the direct emission of the photon from the weak
vertex is mostly CP conserving.  As a result, we must subtract the $DE$
contribution for the determination of $|\eta_{+-}|$.

The $DE$ fraction in $K_L \to \pi^+\pi^-$ has been precisely measured by
E731 \cite{E731_IB} and KTeV \cite{KTEV_IB_1}\cite{KTEV_IB_2}. 
To determine the $DE$ contribution in our $K_{2\pi}$ signal, we
generated three million $K_L \to \pi^+\pi^-\gamma\,(DE)$ decays and
applied the \pip\pim~selection cuts. The acceptance for this
decay mode is only $\sim7\%$ because of the hard
photon spectrum, and the effective $DE$ fraction is: $DE = (0.19 \pm
0.01)\,\%$. Subtracting the $DE$ contribution from the signal, 
we obtain for the \ktpi~branching ratio including 
$K_L \to \pi^+\pi^-\gamma(IB)$:
$$
BR(K_L \to \pi^+\pi^- + \pi^+\pi^-\gamma(IB)) = (1.941 \pm 0.019) 
\times 10^{-3}\,,
$$
with $BR(K_L \to \pi e \nu) = 0.4022 \pm 0.0031$, which was taken from
the NA48 measurement \cite{na48_vus}, but updated for the following
reason: in this experiment, the external error from the
\kltontpi~normalization was the main source of experimental
uncertainty. However, since then the experimental situation in the knowledge of
the branching ratio \kltontpi~has improved, and we recalculated
$BR(K_L \to \pi e \nu)$ with a new $3\pi^0$ normalization, using a
weighted mean of the KTeV \cite{KTEV_eta+-} and the KLOE
\cite{KLOE_kl_decays} results: $BR(3 \pi^0) = (19.68 \pm 0.26)\,\%$
(with enlarged error following PDG rules).
%
%
\subsection{The CP Violation Parameter \etapm}
Using our result for $BR(\kltocpipi)$,  we finally determine the
CP violation parameter
$$
|\eta_{+-}| = \sqrt{\frac{BR(K_L \to \pi^+\pi^-)}{BR(K_S \to
    \pi^+\pi^-)} \cdot \frac{\tau_{KS}}{\tau_{KL}} } = (2.223 \pm
0.012) \times 10^{-3}\,,
$$
taking as further input values the most precise single measurements:
\begin{tabbing}
-\quad\= $\tau_{KS} = (0.89598 \pm 0.00070)\times 10^{-10}$\,s\qquad\quad\= NA48
\cite{na48_ks_lifetime}\,,\\[0.8ex]
-\> $\tau_{KL} = (5.084 \pm 0.023)\times 10^{-8}$\,s\> KLOE 
\cite{KLOE_kl_decays}\,,\\[0.8ex]
-\> $BR(K_S \to \pi^+\pi^-) = 0.69196 \pm 0.00051$\> KLOE \cite{KLOE_ks_pipi}\,.
\end{tabbing}
%
%
%
\subsection{Comparison of Results}
In \Fig{compare_results_etapm}, we compare our results for
\ratio~(left) and \etapm~(right) with the measurements by KTeV
\cite{KTEV_eta+-} and the PDG 2004 values \cite{PDG}. For \etapm, the
new KLOE result \cite{KLOE_kl_pipi} is shown, too.
Using the same values for \kl~lifetime \cite{KLOE_kl_decays} and
$BR(K_S \to \pi^+\pi^-)$ \cite{KLOE_ks_pipi} to determine \etapm, the
results from KLOE and NA48 are correlated. The measurements performed
by the three experiments jointly contradict the former PDG values.  

For the corresponding comparison of $BR(\kltocpipi)$ results in
\Fig{compare_results_BRk2pi}, it is important to point out the
treatment of the radiative decays. The contribution of direct
emission is claimed to be negligible in the KTeV result. The KLOE 
measurement, however, is fully inclusive with respect to final-state
radiation, including both the inner bremsstrahlung and the
(CP conserving) direct emission components. Adding the $DE$
component, our value is in very good agreement with the KLOE
result.
\begin{figure}[ht]
\begin{center}
\includegraphics[width=13.8cm]{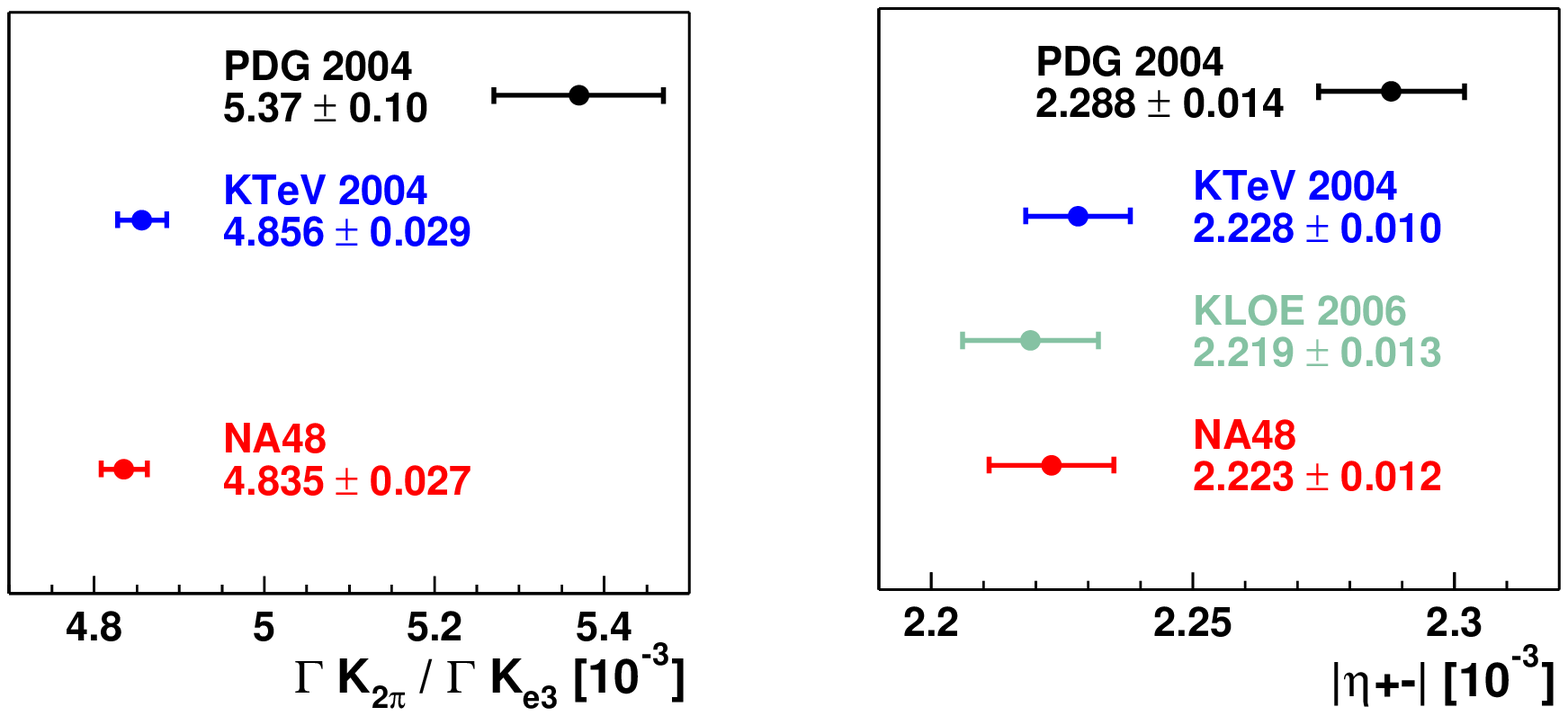}
\parbox{12cm}{\caption[] {\labf{compare_results_etapm}Comparison of
    results for \ratio~(left) and \etapm~(right). The date
    represents the year of publication.}}
\end{center}
\end{figure}
\begin{figure}[ht]
\begin{center}
\includegraphics[width=13.8cm]{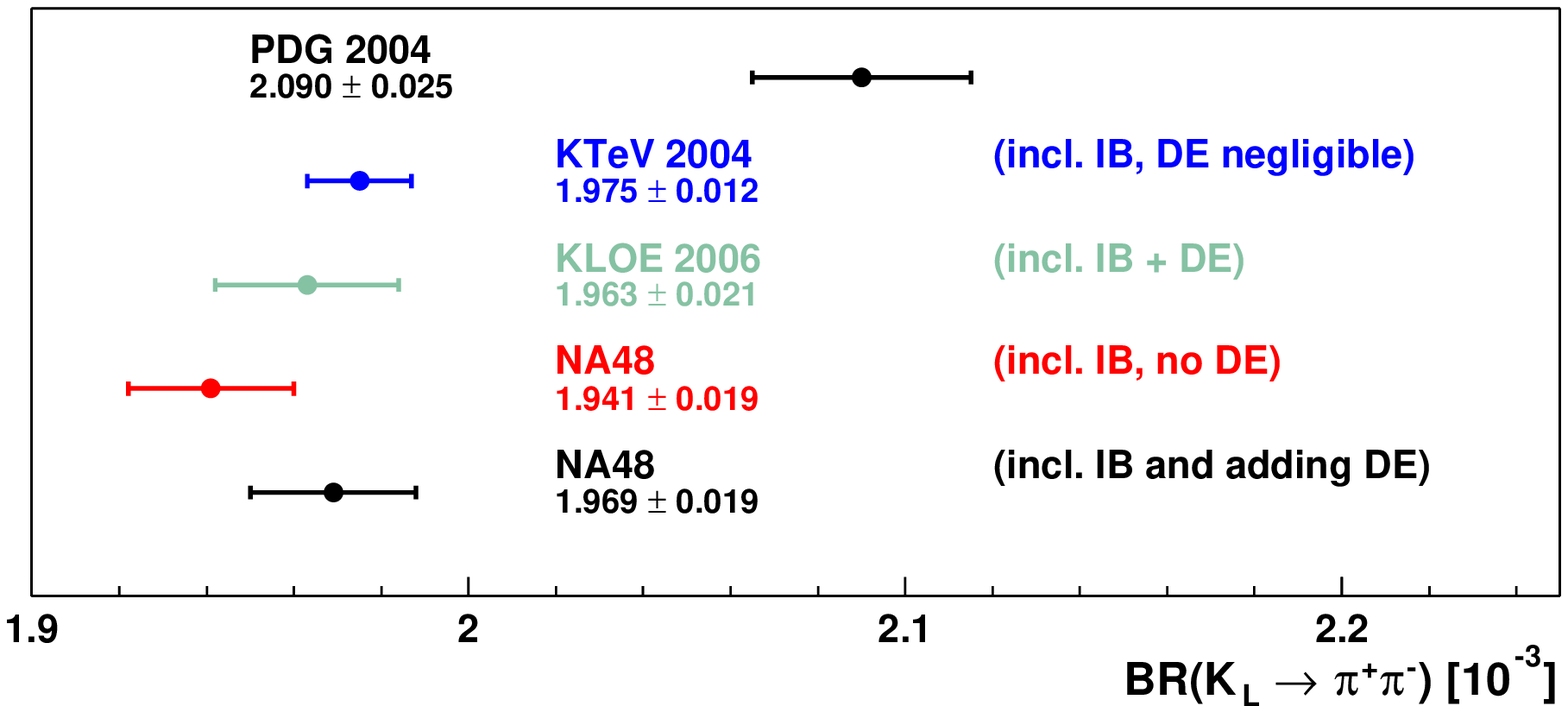}
\parbox{12cm}{\caption[] {\labf{compare_results_BRk2pi}Comparison of
    results for $BR(\kltocpipi)$.}} 
\end{center}
\end{figure}
%
%
\subsection{Conclusions}
In summary, we performed a new, precise measurement of the ratio
\ratio, and extracted results for the \kltocpipi~branching ratio and
the CP violation parameter \etapm. All results contradict the values
reported by the PDG \cite{PDG}, but are in good agreement with recent
measurements obtained by KTeV \cite{KTEV_eta+-} and KLOE \cite{KLOE_kl_pipi}.
\section{Acknowledgements}

We gratefully acknowledge the continuing support of the technical
staff of the participating institutes and their computing centers.

%

\end{document}

%% file: authorlist.tex
\collab{NA48 Collaboration}
\author{A.~Lai},
\author{D.~Marras}
\address{Dipartimento di Fisica dell'Universit\`a e Sezione dell'INFN di Cagliari, \\ I-09100 Cagliari, Italy} 
\author{A.~Bevan},
\author{R.S.~Dosanjh\thanksref{threfRAL1}},
\author{T.J.~Gershon\thanksref{threfRAL2}},
\author{B.~Hay},
\author{G.E.~Kalmus},
\author{C.~Lazzeroni},
\author{D.J.~Munday},
\author{E.~Olaiya\thanksref{threfRAL}},
\author{M.A.~Parker},
\author{T.O.~White},
\author{S.A.~Wotton}
\address{Cavendish Laboratory, University of Cambridge, Cambridge, CB3~0HE, U.K.\thanksref{thref3}}
\thanks[thref3]{Funded by the U.K.\ Particle Physics and Astronomy Research Council}
\thanks[threfRAL1]{Present address: Ottawa-Carleton Institute for Physics, Carleton University, Ottawa, Ontario K1S 5B6, Canada}
\thanks[threfRAL2]{Present address: High Energy Accelerator Research Organization (KEK), Tsukuba, Japan}
\thanks[threfRAL]{Present address: Rutherford Appleton Laboratory, Chilton, Didcot, Oxon, OX11~0QX, U.K.}

\author{G.~Barr\thanksref{threfZX1}},
\author{G.~Bocquet},
\author{A.~Ceccucci},
\author{T.~Cuhadar-D\"onszelmann\thanksref{threfZX2}},
\author{D.~Cundy\thanksref{threfZX}},
\author{G.~D'Agostini},
\author{N.~Doble\thanksref{threfPisa}},
\author{V.~Falaleev},
\author{L.~Gatignon},
\author{A.~Gonidec},
\author{B.~Gorini},
\author{G.~Govi},
\author{P.~Grafstr\"om},
\author{W.~Kubischta},
\author{A.~Lacourt},
\author{A.~Norton},
\author{S.~Palestini},
\author{B.~Panzer-Steindel},
\author{H.~Taureg},
\author{M.~Velasco\thanksref{threfNW}},
\author{H.~Wahl\thanksref{threfHW}}
\address{CERN, CH-1211 Gen\`eve 23, Switzerland}
\thanks[threfZX1]{Present address: Department of Physics, University of Oxford, Denis Wilkinson Building, Keble Road, Oxford, UK, OX1 3RH}
\thanks[threfZX2]{Present address: University of British Columbia, Vancouver, BC, Canada, V6T 1Z1} 
\thanks[threfZX]{Present address: Istituto di Cosmogeofisica del CNR di Torino, I-10133~Torino, Italy}
\thanks[threfPisa]{Present address: Scuola Normale Superiore e Sezione dell'INFN di Pisa, I-56100~Pisa, Italy}
\thanks[threfNW]{Present address: Northwestern University, Department of Physics and Astronomy, Evanston, IL~60208, USA}
\thanks[threfHW]{Present address: Dipartimento di Fisica dell'Universit\`a e Sezione dell'INFN di Ferrara, I-44100~Ferrara, Italy}

\author{C.~Cheshkov\thanksref{threfCERN}},
\author{A.~Gaponenko},
\author{P.~Hristov\thanksref{threfCERN}},
\author{V.~Kekelidze},
\author{L.~Litov},
\author{D.~Madigozhin},
\author{N.~Molokanova},
\author{Yu.~Potrebenikov},
\author{S.~Stoynev\thanksref{threfNW}},
\author{G.~Tatishvili\thanksref{threfCM}},
\author{A.~Tkatchev},
\author{A.~Zinchenko}
\address{Joint Institute for Nuclear Research, Dubna, 141980, Russian Federation}  
\thanks[threfCERN]{Present address: CERN, CH-1211 Geneva~23, Switzerland}
\thanks[threfCM]{Present address: Carnegie Mellon University, Pittsburgh, PA~15213, USA}
\author{I.~Knowles},
\author{V.~Martin\thanksref{threfNW}},
\author{R.~Sacco\thanksref{threfSacco}},
\author{A.~Walker}
\address{Department of Physics and Astronomy, University of Edinburgh, JCMB King's Buildings, Mayfield Road, Edinburgh, EH9~3JZ, U.K.} 
\thanks[threfSacco]{Present address: Department of Physics, Queen Mary, University of London, Mile End Road, London, E1 4NS}
\newpage
\author{M.~Contalbrigo},
\author{P.~Dalpiaz},
\author{J.~Duclos},
\author{P.L.~Frabetti\thanksref{threfFrabetti}},
\author{A.~Gianoli},
\author{M.~Martini},
\author{F.~Petrucci},
\author{M.~Savri\'e}
\address{Dipartimento di Fisica dell'Universit\`a e Sezione dell'INFN di Ferrara, I-44100 Ferrara, Italy}
\thanks[threfFrabetti]{Present address: Joint Institute for Nuclear Research, Dubna, 141980, Russian Federation}
\author{A.~Bizzeti\thanksref{threfXX}},
\author{M.~Calvetti},
\author{G.~Collazuol\thanksref{threfPisa}},
\author{G.~Graziani},
\author{E.~Iacopini},
\author{M.~Lenti},
\author{F.~Martelli\thanksref{thref7}},
\author{M.~Veltri\thanksref{thref7}}
\address{Dipartimento di Fisica dell'Universit\`a e Sezione dell'INFN di Firenze, I-50125~Firenze, Italy}
\thanks[threfXX]{Dipartimento di Fisica dell'Universit\`a di Modena e Reggio Emilia, I-41100~Modena, Italy}
\thanks[thref7]{Istituto di Fisica dell'Universit\`a di Urbino, I-61029~Urbino, Italy}
\author{H.G.~Becker},
\author{K.~Eppard},
\author{M.~Eppard\thanksref{threfCERN}},
\author{H.~Fox\thanksref{threfFB}},
\author{A.~Kalter},
\author{K.~Kleinknecht},
\author{U.~Koch},
\author{L.~K\"opke},
\author{P.~Lopes da Silva}, 
\author{P.~Marouelli},
\author{I.~Pellmann\thanksref{threfDESY}},
\author{A.~Peters\thanksref{threfCERN}},
\author{B.~Renk},
\author{S.A.~Schmidt},
\author{V.~Sch\"onharting},
\author{Y.~Schu\'e},
\author{R.~Wanke},
\author{A.~Winhart},
\author{M.~Wittgen\thanksref{threfSLAC}}
\address{Institut f\"ur Physik, Universit\"at Mainz, D-55099~Mainz, Germany\thanksref{thref6}}
\thanks[thref6]{Funded by the German Federal Minister for Research and Technology (BMBF) under contract 7MZ18P(4)-TP2}
\thanks[threfDESY]{Present address: DESY Hamburg, D-22607~Hamburg, Germany}
\thanks[threfFB]{Present address: Physikalisches Institut, D-79104~Freiburg, Germany}
\corauth[cor]{Corresponding author.\\{\em Email address:} burkhard.renk@uni-mainz.de}
\thanks[threfSLAC]{Present address: SLAC, Stanford, CA~94025, USA}
\author{J.C.~Chollet},
\author{L.~Fayard},
\author{L.~Iconomidou-Fayard},
\author{J.~Ocariz},
\author{G.~Unal\thanksref{threfCERN}},
\author{I.~Wingerter-Seez}
\address{Laboratoire de l'Acc\'el\'erateur Lin\'eaire, IN2P3-CNRS,Universit\'e de Paris-Sud, 91898 Orsay, France\thanksref{threfOrsay}}
\thanks[threfOrsay]{Funded by Institut National de Physique des Particules et de Physique Nucl\'eaire (IN2P3), France}
\author{G.~Anzivino},
\author{P.~Cenci},
\author{E.~Imbergamo},
\author{P.~Lubrano},
\author{A.~Mestvirishvili},
\author{A.~Nappi},
\author{M.~Pepe},
\author{M.~Piccini}
\address{Dipartimento di Fisica dell'Universit\`a e Sezione dell'INFN di Perugia, \\ I-06100 Perugia, Italy}
\author{L.~Bertanza},
\author{R.~Carosi},
\author{C.~Cerri},
\author{M.~Cirilli\thanksref{threfCERN}},
\author{F.~Costantini},
\author{R.~Fantechi},
\author{S.~Giudici},
\author{I.~Mannelli},
\author{G.~Pierazzini},
\author{M.~Sozzi}
\address{Dipartimento di Fisica, Scuola Normale Superiore e Sezione dell'INFN di Pisa, \\ I-56100~Pisa, Italy} 
\author{J.B.~Cheze},
\author{J.~Cogan},
\author{M.~De Beer},
\author{P.~Debu},
\author{A.~Formica},
\author{R.~Granier de Cassagnac\thanksref{threfEcolePoly}},
\author{E.~Mazzucato},
\author{B.~Peyaud},
\author{R.~Turlay\thanksref{threfDECEASED}},
\author{B.~Vallage}
\address{DSM/DAPNIA - CEA Saclay, F-91191 Gif-sur-Yvette, France} 
\thanks[threfEcolePoly]{Present address: Laboratoire Leprince-Ringuet,
\'Ecole polytechnique (IN2P3, Palaiseau, 91128 France}
\thanks[threfDECEASED]{Deceased}

%
%
%
\author{M.~Holder},
\author{A.~Maier\thanksref{threfCERN}},
\author{M.~Ziolkowski}
\address{Fachbereich Physik, Universit\"at Siegen, D-57068 Siegen, Germany\thanksref{thref8}}
\thanks[thref8]{Funded by the German Federal Minister for Research and Technology (BMBF) under contract 056SI74}
\author{R.~Arcidiacono},
\author{C.~Biino},
\author{N.~Cartiglia},
\author{R.~Guida}, 
\author{F.~Marchetto}, 
\author{E.~Menichetti},
\author{N.~Pastrone}
\address{Dipartimento di Fisica Sperimentale dell'Universit\`a e Sezione dell'INFN di Torino, I-10125~Torino, Italy} 
\author{J.~Nassalski},
\author{E.~Rondio},
\author{M.~Szleper\thanksref{threfNW}},
\author{W.~Wislicki},
\author{S.~Wronka}
\address{Soltan Institute for Nuclear Studies, Laboratory for High Energy Physics, PL-00-681~Warsaw, Poland\thanksref{thref9}}
\thanks[thref9]{Supported by the KBN under contract SPUB-M/CERN/P03/DZ210/2000 and using computing resources of the
Interdisciplinary Center for Mathematical and Computational Modelling of the University of Warsaw.}
\author{H.~Dibon},
\author{G.~Fischer},
\author{M.~Jeitler},
\author{M.~Markytan},
\author{I.~Mikulec},
\author{G.~Neuhofer},
\author{M.~Pernicka},
\author{A.~Taurok},
\author{L.~Widhalm}
\address{\"Osterreichische Akademie der Wissenschaften, Institut f\"ur Hochenergiephysik, A-1050~Wien, Austria\thanksref{thref10}}
\thanks[thref10]{Funded by the Federal Ministry of Science and Transportation under the contract GZ~616.360/2-IV GZ 616.363/2-VIII, 
and by the Austrian Science Foundation under contract P08929-PHY.}